\documentclass[longnamesfirst,apjl]{emulateapj}
\usepackage{apjfonts,natbib,epsfig}
\tighten

\newcommand{\beq}{\begin{equation}}
\newcommand{\eeq}{\end{equation}}
\newcommand{\beqa}{\begin{eqnarray}}
\newcommand{\eeqa}{\end{eqnarray}}

\newcommand{\lin}{{\rm lin}}

\begin{document}

\title{Angular Diameter Distance Measurement with Galaxy Clustering in the Multipole Space}

\author{Asantha Cooray}
\affil{Center for Cosmology, Department of Physics and Astronomy, University of California, Irvine, CA 92697}

\righthead{Distance Measurements from Galaxy Clustering}
\lefthead{COORAY}
\begin{abstract}

The shape of the angular power spectrum of galaxies in the linear regime is defined by
the horizon size at the matter-radiation equality. When calibrated by cosmic
microwave background measurements, the shape of the clustering spectrum can be used as a standard ruler to estimate angular diameter distance
as a function of redshift at which galaxy clustering is measured.
We apply the proposed cosmological test of Cooray et al. (2001) to a recent set of 
luminous red galaxy angular clustering spectra from the Sloan Digital Sky Survey  between redshifts of 0.2 and 0.6.
Using the overall shape of the clustering power spectrum in the linear regime, we measure comoving 
angular diameter distances to eight redshift bins by marginalizing over the  bias factors that determine the overall amplitude of the clustering spectrum in each of the bins. The Hubble constant consistent
with these distance estimates is $68.5^{+6.7}_{-6.1}$ km s$^{-1}$ Mpc$^{-1}$
at the 68\% confidence level. We comment on the expected improvements with future surveys and the potential to measure
dark energy parameters with this method.

\end{abstract}

\section{Introduction}
\label{sec:introduction}

The cosmic distance scale, either in terms of the luminosity using standard candles or the size using
standard rulers, is one of the strongest probes of the cosmological expansion rate of the Universe and the dark energy
responsible for recent acceleration (e.g., Huterer \& Turner 2001).
It is well known that features in the clustering spectrum that are associated with known physical scales
provide a basis to establish standard ruler techniques to measure distances between us and the redshift
at which clustering is measured (Eisenstein et al. 1998). Using the sound horizon
at the last scattering surface as the physical scale, this has been successfully applied to 
determine the distance to redshift  $z \sim 10^3$ using the cosmic microwave 
background (CMB)  anisotropy spectrum (Spergel et al. 2003; Spergel et al. 2006).
The angular power spectrum of sources at low redshifts provides a similar test based on the
standard ruler defined by the overall shape of the projected angular 
power spectrum (Cooray et al. 2001).

In the adiabatic cold dark matter model for structure formation,
this standard ruler is the horizon at matter-radiation equality with the absolute physical scale directly calibrated through
CMB anisotropy information.   Similarly, another physical scale  in the source power spectrum that
is now well explored in the literature is the horizon at the end of Compton-drag that results in baryon oscillations 
(Blake \& Glazebrook 2003; Seo \& Eisenstein 2003). 
While the signature of baryon oscillation has now been detected in the galaxy correlation function
at low redshifts with the Sloan Digital Sky Survey (SDSS; Eisenstein et al. 2005; also with 2dF Galaxy Redshift Survey
by Cole et al. 2005), resulting in a 5\% measurement of the absolute distance to $z \sim 0.35$, this ruler has yet to be applied at 
multiple redshifts.  On the other hand, adequate clustering measurements with wide-field surveys such as SDSS now exist to apply 
the standard ruler technique defined by the overall shape of the angular power spectrum to estimate distances out to several bins
in redshift.

Since this standard ruler technique to measure distances is a purely geometric test using the linear power spectrum, it is  insensitive 
to uncertainties on the exact shape of the galaxy power spectrum at non-linear scales. Issues related to non-linear
clustering and corrections related to the conversion from real space to redshift space clustering
are, however, concerns for distance measurements using baryon oscillations (Eisenstein et al. 2006). In galaxy clustering models based on the
halo model (Cooray \& Sheth 2002) or extensions based on  conditional luminosity functions (Cooray 2006) 
galaxy bias is scale independent at large physical scales corresponding to the linear regime. When estimating distances with angular power spectra the bias can be established concurrently, though, if bias is independently known
one can separate the normalization of the power spectrum  to obtain an extra handle on the cosmological parameters
using the growth function of density fluctuations.

In this {\it Letter} we apply the proposed technique of Cooray et al. (2001) to a recent set of
clustering measurements related to luminous red galaxies at redshifts between 0.2 and 0.6 from the Sloan Digital Sky Survey (Padmanabhan et al. 2006).
The clustering measurements in the multipole space are divided to 8 bins in redshifts using photometric redshifts. 
We measure comoving angular diameter distance to each of these redshift bins and combine them to extract the
Hubble parameter today. We also comment on the extent to which distance measurements can be improved in the future with this technique.

This {\it Letter} is organized as following: In the next Section, we will briefly outline the standard ruler
technique involving the angular power spectrum of galaxies and a brief overview of  clustering measurements used for the analysis.
Section~3 describes results from our model fitting and 
a brief discussion of the extent to which this technique can be extended to measure  dark energy parameters. 

\begin{figure}[!t]
\psfig{file=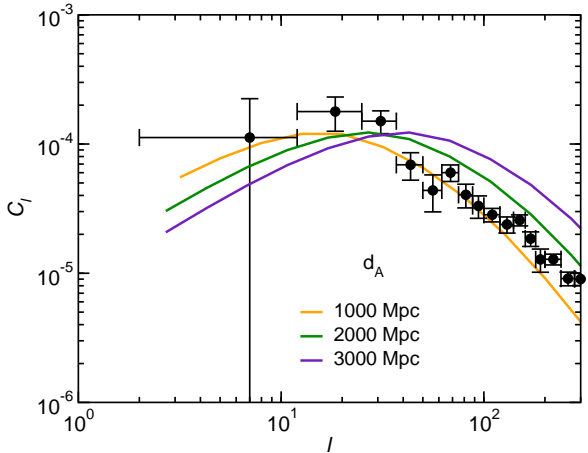,width=2.9in, angle=-90}
\caption{The standard ruler test illustrated using angular power spectrum of luminous red galaxies
in the redshift bin 0.35 to 0.4. The measurements are from Padmanabhan et al. (2006),
while the solid lines show the predictions for clustering based on the linear power spectrum
defined by $\Omega_mh^2=0.132$ and projected at several comoving angular diameter distances. 
In model fitting the data to extract distance, we marginalize over the overall normalization
of the angular spectrum given by the parameter $F(z)$ (equation~4) and the uncertainty in $\Omega_mh^2$ from CMB measurements.
}
\label{fig:cl}
\end{figure}

\begin{figure}[!t]
\psfig{file=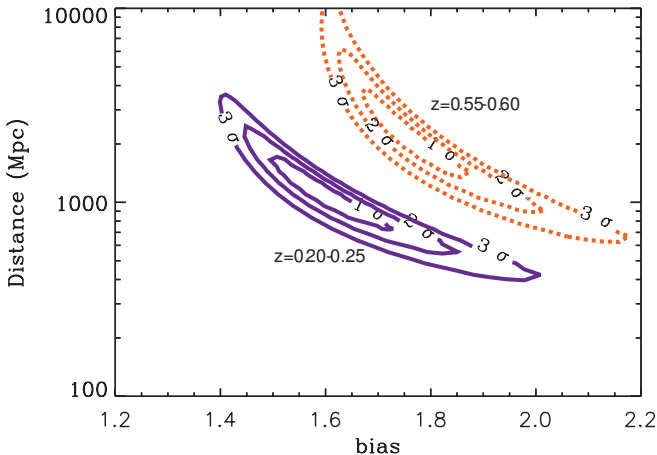,width=3.4in, angle=0}
\caption{The constraints on parameters describing the galaxy bias and the comoving angular diameter distance for two redshift bins at
 $z=0.2$ to 0.25 and $z=0.55$ to 0.6. Note that the two parameters considered for modeling in this paper is the distance and the 
parameter $F(z)$,
though for easy comparison, we have plotted the bias factor by setting other parameters of $F(z)$ in equation~4 with values given by the
recent concordance LCDM model consistent with the WMAP third-year analysis (Spergel et al. 2006).}
\label{fig:chi}
\end{figure}

\begin{figure*}[!t]
\centerline{\psfig{file=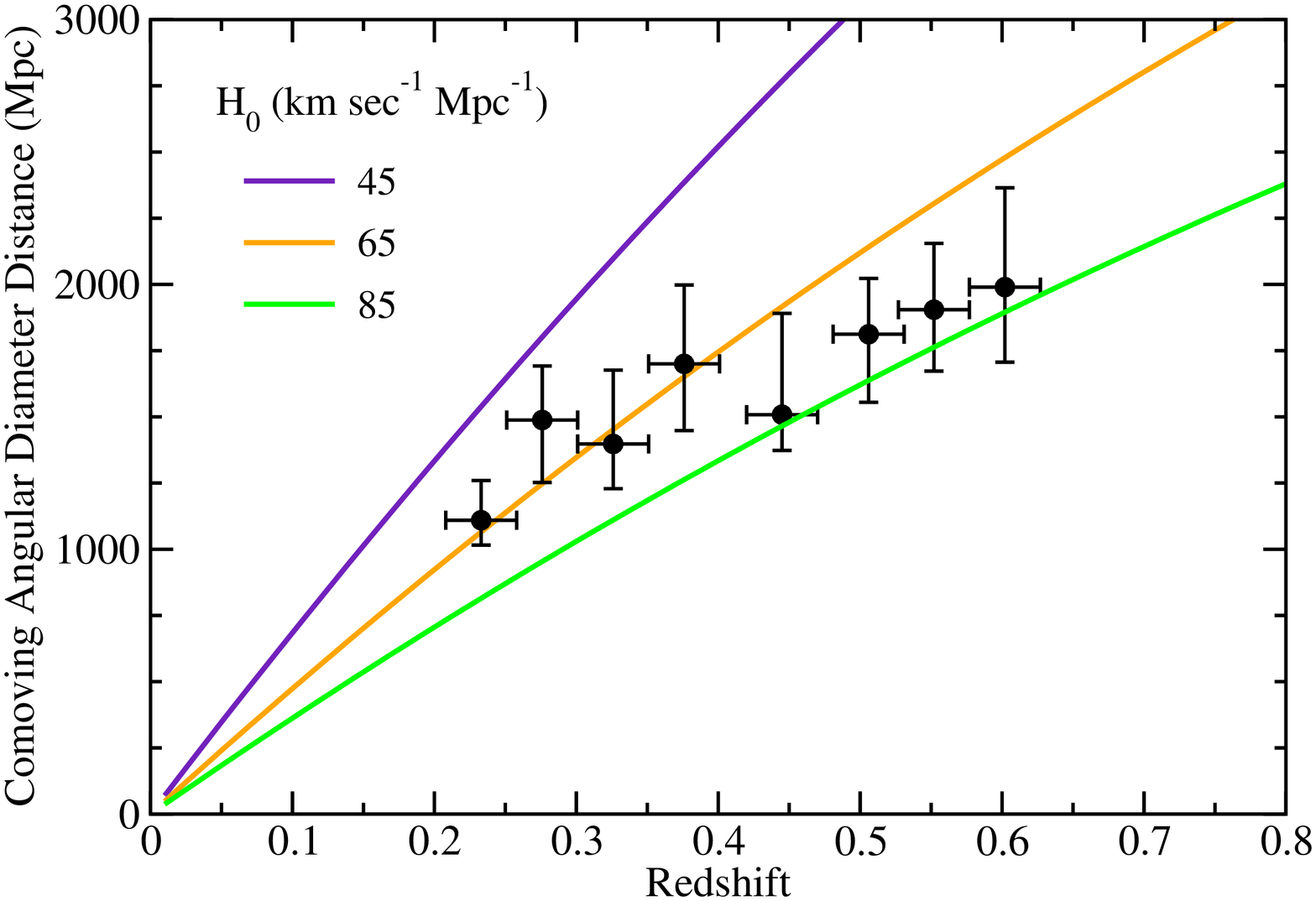,width=2.8in, angle=-90}
\psfig{file=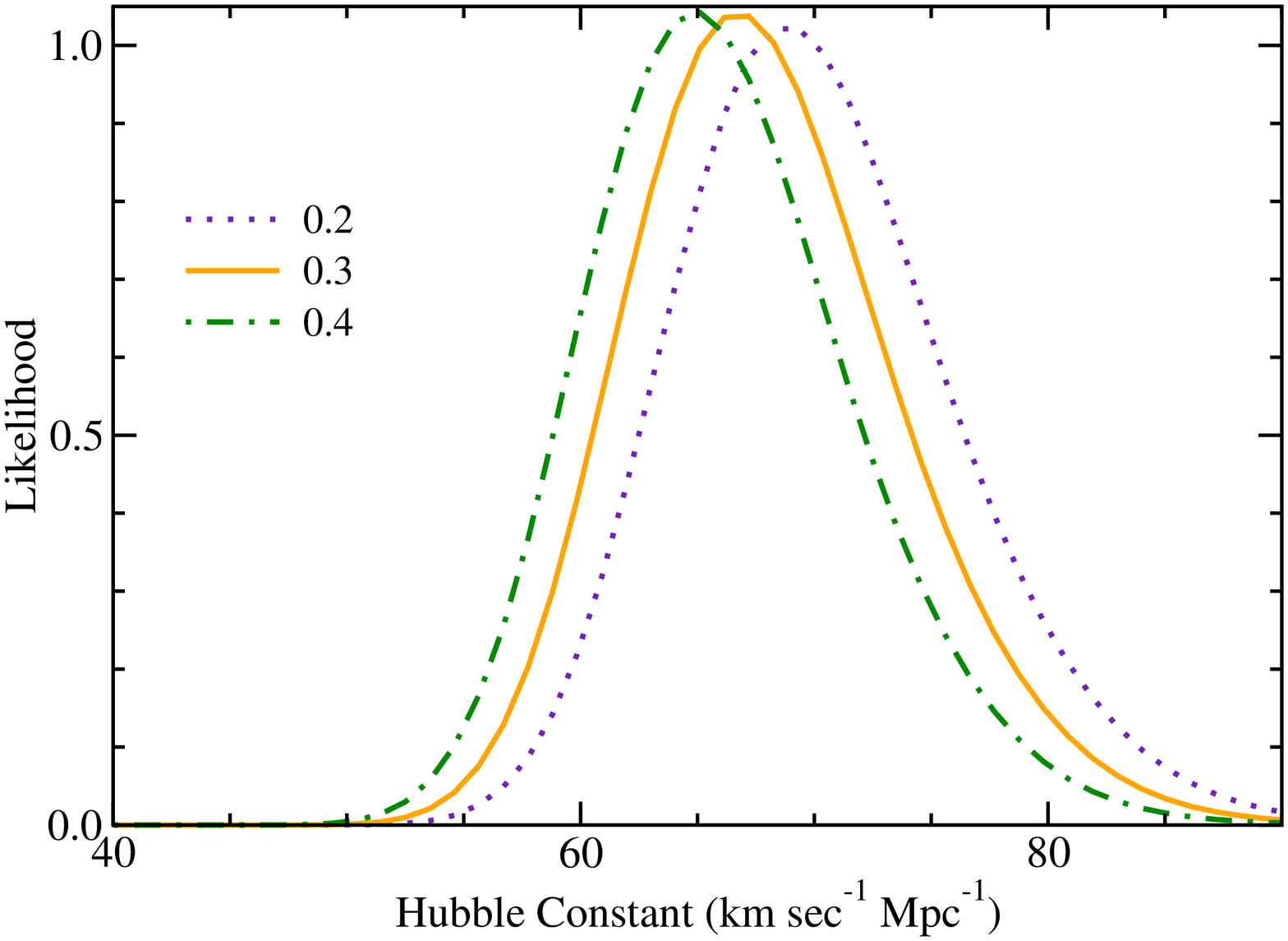,width=2.8in,angle=-90}}
\caption{{\it Left:} The comoving angular diameter distance as  a function of redshift for the eight bins considered
by Padmanabhan et al. (2006) for clustering measurements. The distance estimates are obtained by marginalizing over
the normalization parameter $F(z)$. For reference, we show three difference estimates on the comoving distance
based on the Hubble constant today assuming a flat-$\Lambda$CDM cosmology with $\Omega_m=0.27$.
{\it Right:} The likelihood distribution of the Hubble constant as a function of three assumed values for $\Omega_m$.
For $\Omega_m=0.2$, $H_0=68.5^{+6.7}_{-6.1}$ at the 68\% confidence level.}
\label{fig:h0}
\end{figure*}

\section{Distance Measurement}
\label{sec:distance}

We refer the reader to Cooray et al. (2001) for a detailed discussion related to this
standard ruler method to measure distances. We briefly summarize the main points by first writing the
angular power spectrum of galaxies in $i$th redshift bin through the
Limber (1954; Kaiser 1992) projection of the halo number density power spectrum 
\begin{equation}
C_l^i = \int dz\,
W^2_i(z) {H(z) \over d_A^2(z)} P_{ss}\left(\frac{l}{d_A}; z\right) \,,
\label{eqn:cl}
\end{equation}
where $W_i(z)$ is the normalized redshift distribution of galaxies in the given redshift bin 
so that $\int\, dz\, W_i(z)=1$, $H(z)$ is the Hubble parameter, and $d_A$ is the
angular diameter distance in comoving coordinates\footnote{Note that we use the
comoving coordinates throughout here. There is an additional factor of
$(1+z)$ involved between our definition of $d_A$ and the one
commonly called angular size distance}.

We assume that the sources trace the linear density field with
\begin{equation}
P_{ss}(k;z) = b^2(z) D^2(z) P^\lin(k;0)\,,
\end{equation}
where $b$ is the mass-averaged halo bias parameter, $P^\lin(k;0)$ is
the present day matter power spectrum computed in linear theory, and $D(z)$
is the linear growth function
$\delta^\lin(k;z) = D(z)\delta^\lin(k;0)$.  Note that the halo model suggests
a scale-independent galaxy bias factor in the linear regime (e.g. Cooray \& Sheth 2002).
Equation~(\ref{eqn:cl}) then becomes
\begin{eqnarray}
C_l^i & = & \int dz
 \,W_i^2(z) F(z) P^\lin\left(\frac{l}{d_A^i};0\right) \,, \\
\label{eqn:cz}
F(z) & = & {H(z) \over d_A^2(z)}  D(z)^2 b^2(z)\,.
\label{eqn:fz}
\end{eqnarray}
where the function $F_i(z)$, associated with redshift bin $i$,
contains information on the halo bias, the growth function, the
power spectrum normalization, and terms involved in the 3-D to
2-D projection (such as a $1/d_A^2$ term; see, Eq.~(\ref{eqn:cl})).

The angular diameter distance measurement involves following standard rulers:
the horizon at matter radiation equality, $k_{\rm eq} = \sqrt{2 \Omega_m H_0^2 (1+z_{\rm eq})} \propto \Omega_m h^2$,
which controls the overall shape of the power spectrum, and the sound 
horizon at the end of the Compton drag epoch, $k_{\rm s}(\Omega_m
h^2,\Omega_b h^2)$, which determines baryon acoustic oscillations in the power spectrum.
The angular or multipole locations of these features shift in
redshift as $l_{\rm eq, s} = k_{\rm eq, s} d_A(z_i)$. Thus, the test involves
measuring $C_l^i$ in several redshift bins and using the fact that 
 $l_{\rm eq}$  scales with $d_A(z_i)$ to
constrain the  angular diameter distance as a function of redshift.
Here, to be conservative, we ignore information at non-linear scales and only use measurements in
the linear regime with $l < 200$. We also do not pursue a test involving the baryon features
since these are not detected at a reasonable significance in each of the bins, though, when
clustering measurements in all bins are combined, there is a 2.5 sigma evidence for the presence of oscillations (Padmanabhan et al. 2006).

Note that in our analysis we use clustering measurements in bins $\Delta z$ of 0.05. With such small bins in redshift,
we found the exact shape of  $W_i(z)$ to only lead to few percent corrections to the distance. These
distributions, however, are established with photometric redshift measurements for each of the galaxies in Padmanabhan et al. (2006).
For surveys where exact distributions are less certain, the uncertainty in $W_i(z)$ can be 
marginalized over when estimating distances. 

The standard ruler test assumes that the Limber approximation is adequate for a reliable distance measurement with 
galaxy clustering angular power spectrum.  We have numerically tested the accuracy of the
approximation by integrating the exact formula for $C_l$ and have found the Limber approximation is accurate to 
better than 0.2\% for redshift bins used in the present
analysis. With small redshift bins, there is one complication here involving redshift space distortions associated with peculiar velocities.
Using the Kaiser (1987) formula  related to enhancement of power at large angular scales due to bulk motions, we included terms to related to
velocity-velocity and velocity-density corrections to the galaxy clustering  spectrum (also, see, discussion in
Padmanabhan et al. 2006). These terms mostly affect the overall normalization 
and the determination of galaxy bias parameters and is only responsible for less than a 2\% systematic change to the distance.

\section{Results and Discussion}

Though the technique is clear, what has been lacking so far are adequate clustering measurements that span
large angular scales, or small multipoles, so that the overall shape of the angular power
spectrum of galaxies is determined at several redshifts. The recent measurements of
galaxy clustering in the angular multipole space by Padmanabhan et al. (2006) provide a first 
set of data to apply the proposed technique to estimate distances. The measurements involve the luminous red galaxies
in the Sloan Digital Sky Survey at estimated photometric redshifts between 0.2 and 0.6. The measurements
are subdivided to eight redshift bins of width 0.05. 

In Padmanabhan et al. (2006), the clustering
measurements in each of the redshift bins  and the cross-clustering between redshift bins were
used to estimate the three-dimensional power spectrum of galaxies through an inversion from two dimensional
$C_l$ to three-dimensional $P(k)$. Since no attempt was made to extract individual distances there, other than to
quantify the extent to which a mean distance can be measured for the whole sample, we carry out an
explicit analysis to estimate absolute distances to each of the bins 
following the proposed technique of Cooray et al. (2001).
Figure~1 summarizes the underlying concept behind the proposed method using clustering measurements in one of the redshift
bins as an example. We refer the reader to Padmanabhan et al. (2006) for details on the clustering
measurements as well as a figure of all power spectra in the eight redshift bins.

In addition to distance $d_i(z)$ we also estimate the parameter $F_i(z)$ to each of the redshift bin $i$.
The underlying power spectrum is taken to be the one consistent with WMAP third year analysis 
given in Spergel et al. (2006; WMAP+SDSS). The overall normalization of the
matter power spectrum is not an important parameter for this analysis since it's uncertainty can also
be included with $F_i(z)$. The main parameter of importance is $\Omega_mh^2$ and we marginalize over
the WMAP uncertainty by assuming that the likelihood is a Gaussian. In future, a more thorough
analysis could jointly fit CMB data and angular power spectra in each of the bins at the same time
though we do not pursue such an analysis due to fractionally larger uncertainties in the galaxy clustering
spectra than the CMB clustering spectrum.

In Figure~2, we highlight the constraints on the distances to two redshift bins. For illustrative
purposes and easy comparison with Padmanabhan et al. (2006) results,
instead of the normalization parameter $F_i(z)$, using the $\Lambda$CDM concordance cosmology,
we show the bias factor $b_i(z)$, consistent with each of the estimated $F_i(z)$. The parameter fitting makes use of the full
covariance matrix of the clustering measurements. Our estimates for bias factors are generally
consistent with those given in Padmanabhan et al. (2006) with small differences arising due to
small variations in the concordance parameters considered when estimating $F(z)$ and the description of the  peculiar velocity spectrum
related to redshift space distortion corrections.

In Figure~3 left panel, we marginalize over $F_i(z)$ and plot $d_i(z)$ as a function of redshift to make a
Hubble diagram of comoving angular diameter distances at multiple redshifts using the
standard ruler captured by the angular clustering spectrum of galaxies. To each of the redshift bins,
we measure distances with a typical accuracy of $\sim 10$\% to 20\%. In Figure~3 right panel,
we fit the distance estimates to extract the Hubble constant today. We only fit a single parameter
to the Hubble diagram as we found out that there is no significant information related to, say, dark energy parameters
with distances measured to the accuracies found here. The implied Hubble constant is $68.5^{+6.7}_{-6.1}$ km s$^{-1}$ Mpc$^{-1}$,
at the 68\% confidence level. This value is consistent with recent estimates of the Hubble constant
based on WMAP and large-scale structure power spectrum (Spergel et al. 2006) as well as independent measurements 
involving the Sunyaev-Zel'dovich effect (Bonamente et al. 2005). 

The uncertainty in the estimated Hubble constant is slightly larger than the $\sim$ 6\% 
accuracy of mean distance to a mid redshift of the range suggested in Padmanabhan et al. (2006; see their Section~5.1).
The difference is partly due to our use of only the overall shape for the standard ruler method while
the suggestion in Padmanabhan et al. (2006) also makes use of the implied overall 2.5$\sigma$ detection 
of baryon oscillations in the combined power spectrum from angular spectra in each of the 8 redshift bins.  
The accuracy of $\sim$ 10\% for the Hubble constant we find  here, however,
is in good agreement with the suggestion in Padmanabhan et al. (2006) that, with a negligible baryon fraction, a mean distance to
$z\sim 0.5$ can be established with an accuracy of 10\% using the combined clustering power spectrum.
Here, we have explicitly carried out the necessary analysis and have established
a Hubble diagram using clustering measurements for the first time using a standard ruler technique.

While the distances are estimated roughly to fractional accuracies between 10\% to 20\% in most redshift bins,
for precision dark energy measurements, we find that distance estimates with accuracies below a few percent are required.
Using the overall shape of the clustering spectrum alone, such measurements are
challenging, though not impossible. The clustering measurements used for the present analysis involve
on average 40,000 to 60,000 galaxies in each of the redshift bins over a total imaging area of 3500 square degrees.
For 5\% distance measurements in 0.05 redshift bins, the sample sizes of galaxies in each of the redshift bins must be increased
by a factor more than 4. This could be achieved by increasing the sky survey area to roughly a half of the sky for galaxy samples
out to a  redshift of 0.5 or so. Though this is a large requirement, such data will be available with expected
catalogs  from survey instruments such as the Large Synoptic Survey Telescope (LSST; Zhan et al. 2006).

Similarly, one needs to push these measurements beyond redshift of 0.6 now probed with SDSS.
At high redshifts,  the area requirement becomes smaller as the volume is increased. We find that
adequate distance measurements out to $z \sim 1 to 1.5$ can be achieved with surveys
of a few thousand square degrees provided that galaxy samples are still selected to be
those involving large bias factors so the clustering amplitude is large. 
The proposed Dark Energy Survey (DES; Annis et al. 2005)  will provide a first dataset to measure distances with this standard
ruler technique out to redshifts of 1 and slightly larger. 
At redshifts between 2 and 4, a very large area ($> 100$ square degrees) Lyman-break galaxy survey
could provide adequate measurements for 10\% accurate absolute distances to such redshifts.

In any case, it is unlikely that precision dark energy measurements can be achieved
with distances measured from the overall clustering shape alone. A factor of 4 to 5 improvement in distance measurements
could be easily achieved with precision detection of baryon oscillations since this feature
is sharper than the physical ruler related to the horizon at matter-radiation equality. Such measurements, however,
must overcome uncertainties and systematics related to non-linear clustering though most concerns have now
shown to lead to at most a few percent systematic corrections to, say, the location of the baryon peak 
(Eisenstein et al. 2006). The angular diameter distances from galaxy clustering 
may be more useful in combination with other cosmological distance measurements, such as luminosity distances from type Ia supernovae
since the combination breaks degeneracies between certain cosmological parameters.
Furthermore, is certain non-standard cosmologies (e.g., Uzan et al. 2004), the 
luminosity and angular diameter distances may not follow the 
usual cosmological relation of $d_L=d_A(1+z)^2$. Thus, in addition to large samples of supernovae, techniques to improve
angular diameter distances measurements must also be pursued in the future for a variety of purposes. 
Thus, it is likely that the standard ruler method based on features in the power spectrum, either the overall shape,
baryon oscillation or both, will remain to be useful.

\acknowledgements 

We thank Nikhil Padmanabhan for making available to us electronic data files of clustering
measurements from Padmanabhan et al. (2006).  This work was completed during author's stay at the Aspen Center for
Physics.

\end{document}